\documentclass[reprint,superscriptaddress,amsmath,amssymb,aps,pra]{revtex4-1}

\usepackage{standalone}
\usepackage[dvipsnames]{xcolor}
\usepackage{graphicx}
\usepackage{dcolumn}
\usepackage{bm}
\usepackage[colorlinks]{hyperref}
\usepackage{pgfplots}

\usepackage{tikz}
\usepackage[utf8]{inputenc}
\usepackage{physics}
\usepackage{amsmath}
\usepackage{braket}
\usepackage{filecontents}

\usepackage{dutchcal}
\usepackage{textcomp}
\usepackage{relsize}

\usepackage[colorlinks]{hyperref}
\newcommand\myshade{80}
\colorlet{mylinkcolor}{ForestGreen}
\colorlet{mycitecolor}{Red}
\colorlet{myurlcolor}{RoyalBlue}
\usepackage[dvipsnames]{xcolor}

\hypersetup{
  linkcolor  = mylinkcolor!\myshade!black,
  citecolor  = mycitecolor!\myshade!black,
  urlcolor   = myurlcolor!\myshade!black,
  colorlinks = true
}

\newcommand{\LPTWON}{
LP2N, Laboratoire Photonique, Num\'erique et Nanosciences,\\
Universit\'e Bordeaux-IOGS-CNRS:UMR 5298, F-33400 Talence, France}

\newcommand{\SOTON}{
School  of  Physics  \&  Astronomy,  University  of  Southampton, \\
Highfield,  Southampton  SO17  1BJ,  United  Kingdom}

\newcommand{\INO}{
Istituto Nazionale di Ottica, INO-CNR, Sesto Fiorentino, 50019, Italy}

\newcommand{\Unibo}{
Dipartimento di Fisica e Astronomia,
Università di Bologna, Bologna, 40127, Italy}

\definecolor{purplemunsell}{rgb}{0.62, 0.0, 0.77}

\begin{document}

\preprint{APS/123-QED}

\title{Loading and Cooling in an Optical Trap via Hyperfine Dark States}

\author{D. S. Naik}
\email{devang.naik@institutoptique.fr}
\affiliation{\LPTWON}

\author{H. Eneriz-Imaz}
\affiliation{\LPTWON}

\author{M. Carey}
\affiliation{\LPTWON}
\affiliation{\SOTON}

\author{T. Freegarde}
\affiliation{\SOTON}

\author{F. Minardi}
\affiliation{\INO}
\affiliation{\Unibo}

\author{B. Battelier}
\affiliation{\LPTWON}

\author{P. Bouyer}
\affiliation{\LPTWON}

\author{A. Bertoldi}
\email{andrea.bertoldi@institutoptique.fr}
\affiliation{\LPTWON}

\date{\today}

\begin{abstract}
We present a novel optical cooling scheme that relies on hyperfine dark states to enhance loading and cooling atoms inside deep optical dipole traps.
We demonstrate a seven-fold increase in the number of atoms loaded in the conservative potential with strongly shifted excited states.
In addition, we use the energy selective dark-state to efficiently cool the atoms trapped inside the conservative potential rapidly and without losses.
Our findings open the door to optically assisted cooling of trapped atoms and molecules which lack the closed cycling transitions normally needed to achieve low temperatures and the high initial densities required for evaporative cooling.
\end{abstract}

\pacs{Valid PACS appear here}

\maketitle

Ultra-cold quantum gases have attracted much attention in recent decades as versatile platforms for investigating strongly correlated quantum systems \cite{Bloch2008} and as the basis for a new class of quantum technologies based on atomic interferometry \cite{Cronin2009,Ludlow2015}. Cooling of an atomic gas to the required temperatures requires a multi-stage process: laser cooling in a magneto-optical trap (MOT); sub-Doppler cooling; loading into a conservative magnetic or optical trap; evaporative cooling. Although quite efficient, this process is only possible for a small subset of alkali and alkali-earth-like atoms that can be initially cooled to low temperature by optical means.

The sub-Doppler cooling phase typically uses light red detuned from the $F \rightarrow F'=F+1$ cycling transition of a D$_2$ line $n S_{1/2} \rightarrow  n P_{3/2}$ \cite{Dalibard1989}, and can be understood in terms of the ``Sisyphus effect'' \cite{Weidemuller1994,Grier2013,Rosi2018,Jarvis2018}.
Sub-Doppler cooling schemes involving dark states (DSs) \cite{FinkelsteinShapiro2019} have emerged as a powerful alternative; they are known as \textit{gray molasses}. Very recently they have been pivotal in obtaining an all-optical BEC in microgravity \cite{Condon2019} and a degenerate Fermi gas of polar molecules \cite{DeMarco2019}. The DSs are coherent superpositions of internal and external (momentum) states that are decoupled from the optical field; their creation does not require cycling $F \rightarrow F'=F+1$ transitions, but can rely on any transitions of the $F \rightarrow F'\leq F$ form.

To prevent the expansion of the cold atom cloud during cooling, it is tempting to combine DS sub-Doppler cooling with spatial confinement in a far-off-resonance optical dipole trap (FORT). However, the DSs are then strongly modified by the trap potential, which typically shortens their lifetime and eventually couples them to the light field.

In this letter, we show that DS cooling can be used in combination with FORT when strong differential light shift are present. We use the effect of FORT trapping light close the $5P_{3/2} \rightarrow 4D_{3/2;5/2}$ transitions of rubidium at 1529 nm \cite{Brantut2008,Bertoldi2010} to maximize the cooling action on the surrounding of the trap and at its borders. We observe an order of magnitude improvement in the number of trapped atoms. Additionally, we explore the possibility of cooling the atomic ensemble in the FORT. We achieve a notable temperature reduction of the trapped sample in a few ms and without losing atoms and analyze the limitations of the protocol and the possible workarounds. \medskip

\begin{figure}[b]
%\includestandalone[scale=0.7]{RbLevelScheme_Detuning}
\includegraphics[scale=0.7]{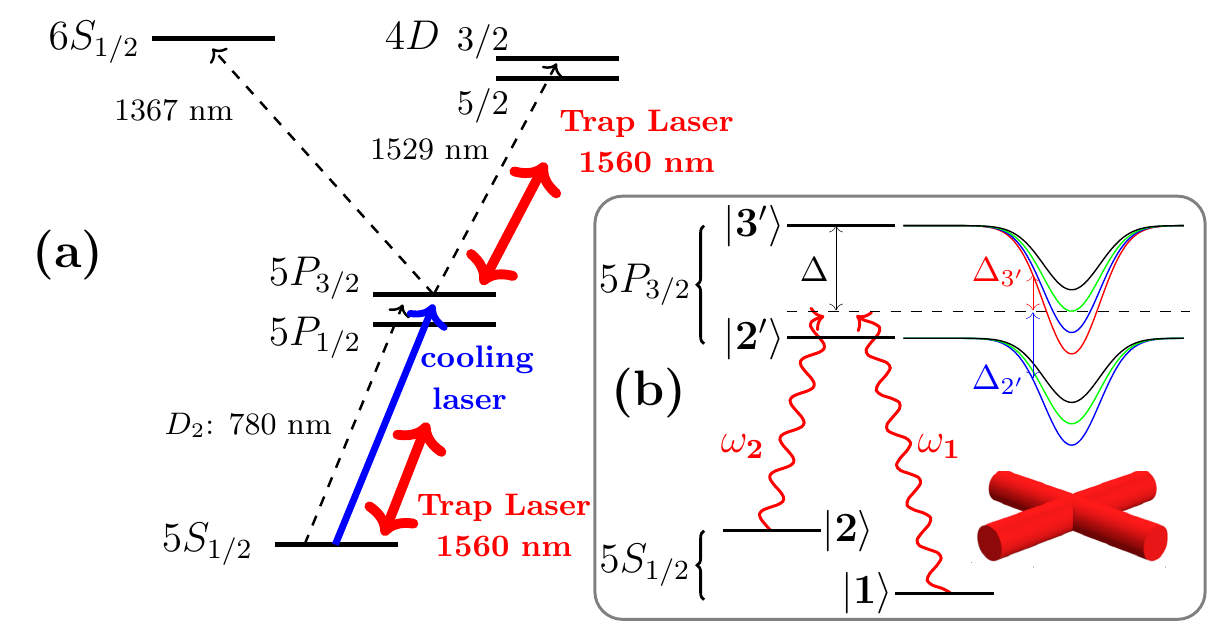}
\caption{\label{fig:LightShiftsSchematic}(a) Relevant energy levels of $^{87}$Rb.
(b) $m_{F'}$-dependent light shifts on the $5P_{3/2}$ levels caused by the FORT at 1560 nm -- not to scale. The red waves show the lasers involved in the Raman scheme, $\Delta_{3'}$ and $\Delta_{2'}$ indicate the cooler detunings at any given point in the trap.
}
\end{figure}

Gray molasses commonly rely on Zeeman dark states (ZDSs), i.e. linear combinations of Zeeman sublevels with different momenta not coupled to lasers blue-detuned with respect to the cooling transition \cite{Papoff1992}. Generally, ZDSs are not eigenstates of the kinetic energy operator, and free evolution turns them into bright, absorbing states. Additionally, the presence of a repumper laser spoils these ZDSs, limiting the cooling efficacy \cite{PhysRevA.51.R2703,PhysRevA.53.1014}.

However, a configuration where the repumper and cooler frequencies are equally detuned from the excited state, i.e. in a Raman configuration (see Fig. \ref{fig:LightShiftsSchematic}(b)), hyperfine dark states (HDSs) can be created. These HDSs have been rigorously studied in the context of coherent population trapping and electromagnetically induced transparency \cite{FinkelsteinShapiro2019,PhysRevLett.47.838,PhysRevA.27.906}, but not yet for gray molasses cooling.

The existence of these HDSs depends on the number of degenerate ground ($N_{g}$) and excited ($N_{e}$) Zeeman states \cite{Papoff1992,FinkelsteinShapiro2019}: (i) for $N_g > N_e$, a DS always exists provided the laser frequencies match the Raman condition; (ii) for $N_g \leq N_e$, additional conditions on the complex Rabi frequencies must be satisfied. For instance, DSs where the connectivity between ground and excited states forms a loop exist if the summed phase of each complex Rabi frequencies is 0 mod($2 \pi$) \cite{Kosachiov1992,FinkelsteinShapiro2019}. These states are particularly relevant for gray molasses cooling because they can be eigenstates of the momentum, thus stable both under free evolution and in presence of a slowly varying external potential.

The DSs for case (ii) can exist in the usual $\sigma_+-\sigma_-$ cooling configuration, provided that that lasers tuned to the cooling $F=2 \rightarrow F'=2$ and repumping $F=1 \rightarrow F'=2$ transitions are both phased locked and retro-reflected to fulfil the phase requirements on the Rabi frequencies. In addition, the excited Zeeman states $\ket{F'=2,m'}$ must be degenerate (see Fig. \ref{fig:LightShiftsSchematic}(b)), e.g. at zero magnetic field or when the connected states experience the same light shifts in the presence of far of resonance light. We used a numerical approach to ascertain the exact DSs for our configuration where light at 1560 nm is used for trapping (see Supplemental Materials \cite{supplMat}). \medskip

Our experimental scheme is described in \cite{Naik2018}. We load and cool $^{87}$Rb atoms at the center of an optical cavity used for the 1560 nm FORT. During the entire MOT stage, the FORT depth is maintained at $\sim$27 $\mu$K. After 2 s of MOT loading, we realize a compressed MOT phase (CMOT) by detuning for 40 ms the MOT beams to $\Delta =-6\, \Gamma$  to the red of the $F=2 \rightarrow F'=3$ atomic resonance and decreasing the optical power by a factor 10 ($\Gamma=2 \pi \cdot 6.066$ MHz is the natural linewidth of the Rb D2 line). Throughout this process, the MOT region overlapping the FORT remains effectively dark due to the large excited state light shifts. The result is a larger density in this region, in a similar fashion to what reported for the dark-SPOT MOT \cite{Ketterle1993}. At this point we begin the sub-Doppler cooling phase. The FORT is ramped up to 170 $\mu$K, simultaneously the cooling beams are detuned to the red in 200 $\mu$s, and we illuminate the atoms for 1 ms. We then turn off the cooling and repumping beams and hold the atoms for 500 ms in the FORT. We finally measure the number of trapped atoms by absorption imaging.

\begin{figure}[t]
%\includestandalone[scale=0.97]{AtomLoading}
\includegraphics[scale=0.97]{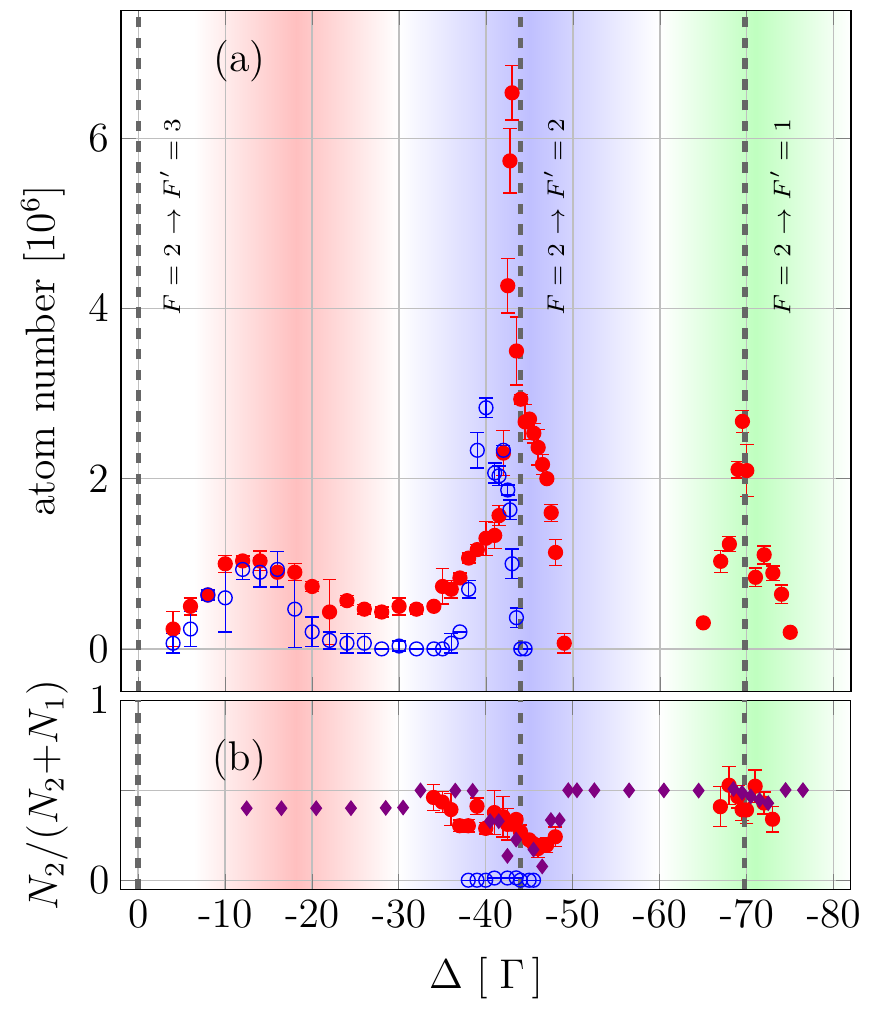}
\caption{\label{fig:AtomLoading} (a) Atoms loaded into the FORT as a function of the cooling laser detuning from the $F = 2 \rightarrow F'= 3$ resonance during the sub-Doppler cooling phase, for an independent repumper (empty blue circles), and for a repumper in Raman condition with the cooler (filled red circles). (b) Population fraction of atoms in $F=2$ with the repumper in the two configurations as above; N$_{1,2}$ indicate the atomic populations of the $F=1,2$ manifolds. The violet points result from a numerical evaluation of the DSs forming at each detuning (see Supplemental Material \cite{supplMat}).}
\end{figure}

Figure \ref{fig:AtomLoading}(a) shows the number of captured atoms as a function of $\Delta$. We used two molasses configurations: first (blue points in Fig. \ref{fig:AtomLoading}(a)) with a repumper at fixed frequency, on-resonance to the $F = 1 \rightarrow F'= 2$ transition; second (red points in Fig. \ref{fig:AtomLoading}(a)) with the repumper phased-locked to the cooling laser by using an EOM at the hyperfine frequency $\omega_{\rm{hf}} = 6.83468$ GHz. This latter configuration leads to the Raman configuration of Fig. \ref{fig:LightShiftsSchematic}(b)). We can identify three regions: (i) $\abs{\Delta}/\Gamma<30$ (red-shaded region), where standard $F = 2 \rightarrow F'= 3$ sub-Doppler cooling takes place, leading to optimum loading at $\abs{\Delta}/\Gamma \simeq 15$ for both configurations; (ii) $30<\abs{\Delta}/\Gamma<60$ (blue-shaded region) and (iii) $60<\abs{\Delta}/\Gamma$ (green-shaded region), where gray molasses occurs, respectively on the $F = 2 \rightarrow F' = 2$ and $F = 2 \rightarrow F' = 1$ transitions.

In the first configuration, only ZDSs can form. Gray molasses leads to a 2.5 times more efficient loading at -40$\, \Gamma$, i.e. blue detuned from the $F=2\to F'=2$ transition. This coincides with an optimum in density when the cooling is applied without the FORT  \cite{Rosi2018}. We find that almost all the atoms are in the $F = 1$ manifold (blue points in Fig. \ref{fig:AtomLoading}(b)) even though the ZDSs are in the $F = 2$ manifold. This can be understood as the effect of the repumper weakly coupling the DSs \cite{PhysRevA.51.R2703,PhysRevA.53.1014}, leading to eventual decay in $F = 1$.

\begin{figure}[t]
%\includestandalone[scale=0.95]{AtomLoadingFortPower}
\includegraphics[scale=0.95]{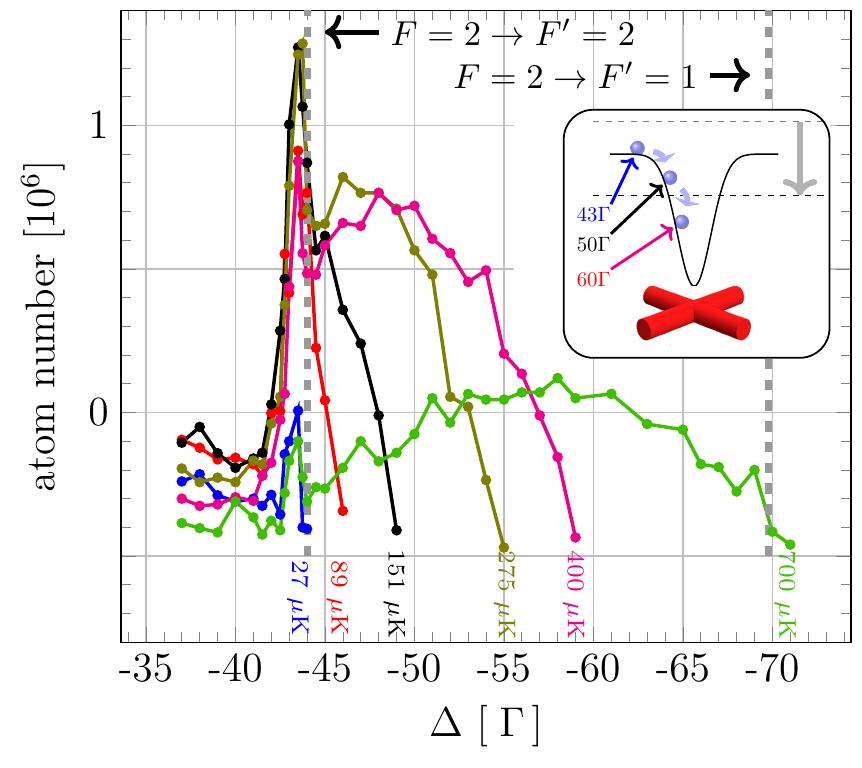}
\caption{\label{fig:AtomLoadingFortPower} HDS cooling near the  $F = 2 \rightarrow F'= 2$ transition at different FORT depths indicated in colors. The sharp peak at -43.5$\, \Gamma$ is given by atoms cooled just outside the FORT. The broader peak underneath corresponds to atoms residing within the FORT volume. Inset: The radially changing FORT power results in a position dependent excited state light shift, exploited to selectively cool the optically trapped atoms at specific radial positions.}
\end{figure}

In the second configuration HDSs can form. We observe a more efficient loading, with an outstanding narrow peak at -43.5$\, \Gamma$, seven times higher than standard red detuned molasses and more than twice higher than ZDS gray molasses. A similar peak is also observed at -69$\, \Gamma$, i.e. blue detuned from the $F = 2 \rightarrow F'= 1$ transition; its lower loading efficiency is partially due to the ramping process of cooler+repumper frequencies from the MOT frequencies, which involves crossing the $F = 2 \rightarrow F'= 2$ transition with associated heating. The atomic population in this region is balanced for all detunings (red points in Fig. \ref{fig:AtomLoading}(b)); all atoms are in balanced HDSs, mediated either by the symmetric excited state $m_{F'} = \pm 1$ or via $m_{F'} = 0$.

We now focus on the sharp peak at -43.5$\, \Gamma$. As in the previous configuration, this coincides with the optimum density reached without FORT-induced light shifts, which is shifted closer to the $F = 2 \rightarrow F'= 2$ transition because of the Raman configuration \cite{Rosi2018}. This suggests that cooling is mostly efficient outside and on the edges of the FORT. This peak sits atop a broader feature which does not abruptly end at the position of the $F = 2 \rightarrow F'= 2$ resonance \footnote{A similar double structure can be resolved in the blue data points of Fig. \ref{fig:AtomLoading}(a), but the peak at -40$\, \Gamma$ is much less intense}. The broader feature shifts and broadens with increasing FORT power while the position of the sharp peak remains constant (Fig. \ref{fig:AtomLoadingFortPower}). The broader feature refers to atoms loaded from within the trapping volume of the FORT; optimal detuning to cool these atoms is larger and position dependent, because of the excited state light shifts imposed by the FORT (Fig. \ref{fig:LightShiftsSchematic}). The shape of the broader loading curve results from a convolution of the atomic density distribution and the spatial FORT profile: the deeper the FORT, the broader the loading curve. It shows that we can spatially address atoms trapped in the FORT by adjusting $\Delta$ \cite{Brantut2008} and suggests a mechanism for further cooling trapped atoms.

Increasing the power of the FORT changes how gray molasses works, which is reflected in the loading curves shown in Fig. \ref{fig:AtomLoadingFortPower}. Two effects could explain our observations. First, the trapping light at 1560 nm induces light shift much larger for the excited state levels than for the ground state ones. Off-resonant scattering from excited state hyperfine levels not involved in the formation of DSs can hinder ZDSs and HDSs, and the resulting decay rate will be strongly related to the energy shifts. This should results in an effective lifetime of the DSs that depends on the local intensity of the trapped light. Second, HDSs relying on two (or more) magnetic sub-levels with different $\left | m_{\rm{F}} \right |$ should be strongly modified and probably inhibited when atoms move in the inhomogeneous FORT intensity. We have performed numerical calculations and found a good agreement between the measured population ratio and the calculated ones (see the violet points in Fig. \ref{fig:AtomLoading}(b) and Supplemental Materials \cite{supplMat}). A detailed experimental and theoretical study would be needed to further confirm these hypothesis, and is beyond the scope of this work.

\begin{figure}
%\includestandalone[scale=0.93]{FreqShiftFORTPower}
\includegraphics[scale=0.93]{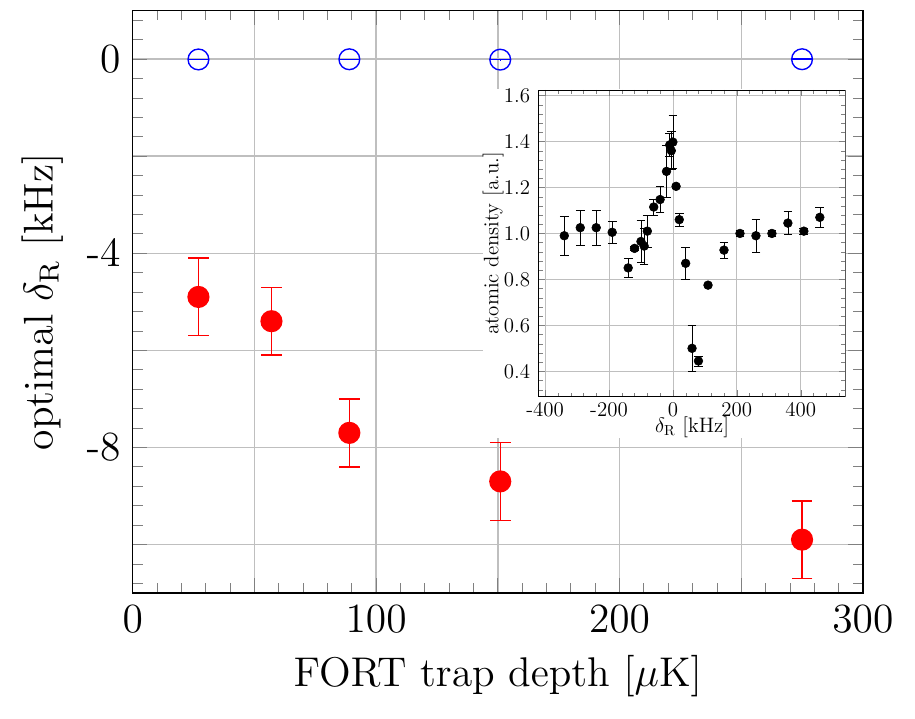}
\caption{\label{fig:FreqShiftFortPower} Raman detuning optimizing the HDS cooling, as a function of the FORT power. The filled red points correspond to HDSs on the $F=2 \rightarrow F'=2$ transition, the empty blue ones to HDSs on the $F=2 \rightarrow F'=1$ transition (where no shift is required to optimize cooling). Inset: HDS cooling efficiency as a function of the Raman detuning frequency, without the FORT.
}
\end{figure}

The deleterious effects of the 1560 nm on the formation of the DSs can be partly offset by tuning the Raman detuning frequency $\delta_{\rm{R}}$. Fig. \ref{fig:FreqShiftFortPower} shows how the optimum value $\delta_{\rm R}$ varies with the FORT power. We measure this value by scanning $\delta_{\rm R}$ to find the maximum peak central density of the atomic cloud. The inset in Fig. \ref{fig:FreqShiftFortPower} shows such a scan for an atomic cloud after 4 ms of gray molasses in free space, without 1560 nm trapping light. The exact Raman condition, $\delta_{\rm{R}} = 0$, gives optimal cooling - that is the lowest temperature and highest central density - and the cooling efficiency falls when the Raman detuning is changed, reaching a minimum at $\delta_{\rm{R}}$=80 kHz. Remarkably, the distance between the maximum and the minimum is much smaller than $\Gamma$ as in similar work with Cs atoms \cite{PhysRevA.98.033419}, but in contrast with Ref. \cite{Grier2013}.

The red points in Fig. \ref{fig:FreqShiftFortPower} refer to the HDS gray molasses on the $F = 2 \rightarrow F'= 2$ transition: an increasing compensation shift is required for higher FORT power. Since for the ground level the vector light shifts is zero, the observed shift must be caused by the effect of the excited state light shifts on the HDSs, however the exact mechanism is still unclear. The blue points refer to the HDS gray molasses on the $F = 2 \rightarrow F'= 1$ transition $\Delta /\Gamma=-69$, for which cooling is optimal at $\delta_{\rm{R}}$ = 0, independent of the FORT power. \medskip

As discussed earlier, Fig. \ref{fig:AtomLoadingFortPower} shows that trapped atoms experience position dependent excited state light shifts, proportional to the local FORT intensity, which allow for addressing spatially these atoms by adjusting $\Delta$. In addition, the position of atoms within a conservative trap is energy dependent, with higher (lower) energy states spending more time near the edges (center) where the light shift is smaller (larger) \cite{Brantut2008}. Therefore at any specific detuning $\Delta$, atoms of a specific energy should be optimally cooled and fall deeper into the FORT. Consequently, if we progressively increase $\Delta$, we should be able to continuously cool the atoms from outside the trap to deep inside it, increasing the density while potentially reaching temperatures close to the recoil limit with no atom loss since our cooling relies on DSs.

\begin{figure}
%\includestandalone[scale=0.75]{temperaturegraph}
\includegraphics[scale=0.75]{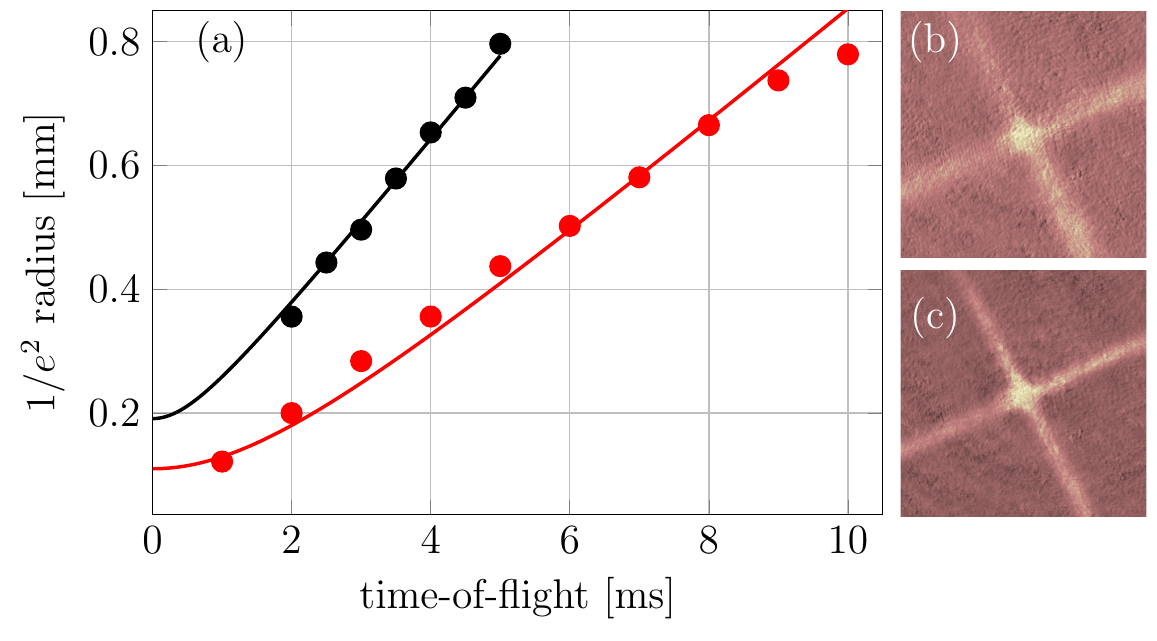}
\caption{\label{fig:temperaturegraph} (a) Ballistic expansion of the atomic cloud released from the FORT: cloud size versus time-of-flight. The fits (solid lines) yield a temperature of 198 $\mu$K before the cooling sweep (black points) and 48 $\mu$K after it (red points).  Absorption images of the atoms before (b) and after (c) the cooling sweep, taken with a time-of-flight of 700 $\mu$s.}
\end{figure}

To investigate this, we adiabatically increase the FORT depth to $\sim$1 mK after loading, to increase the excited state light shifts and, thereby, the energy selectivity. We sweep the detuning in 6 ms from -55$\, \Gamma$ to $-68\, \Gamma$ and observe a temperature reduction from 198 $\mu$K to 48 $\mu$K (Fig. \ref{fig:temperaturegraph}(a)). A slower sweep does not lower the final temperature, whereas increasing the final detuning beyond $-68\, \Gamma$ results in a complete atom loss as the laser frequency becomes resonant with the $F = 2 \rightarrow F'= 2$ transition at the center of the FORT.
Higher energy sensitivity can be achieved by further increasing the FORT depth, as shown in Fig. \ref{fig:AtomLoadingFortPower}. However, for large shifts, the $F = 2 \rightarrow F' = 3$, 267 MHz higher than the $F = 2 \rightarrow F' = 2$ transition, will become resonant at the center of the trap, thus hindering  formation of HDSs \cite{Papoff1992} and eventually causing heating. This main limitation to our scheme points to a straightforward remedy: exploit the $5S_{1/2} \rightarrow 5P_{1/2}$ D1 transition, where no higher $F' = 3$ level exists. This solution should permit recoil temperatures in timescales of milliseconds without loss of atoms. \medskip

To conclude, we have demonstrated the potential of a new type of all-optical loading and cooling in a far-off-resonance trap using gray molasses cooling relying on hyperfine dark states, combined with large light shifts of the excited levels. The method could open new avenues in the production of ultra-cold atomic and molecular gases thanks to many key features: high speed, minimal scattering, no atom loss and no need for cycling transitions. Remarkably, cooling and loading happen simultaneously, which solves the mode matching issue for optical dipole traps preventing the need for an intermediate step often relying on magnetic trapping. The cooling scheme looks optimal for the rapid production of ultra-cold gases in unusual geometries \cite{Xin2018,Xin2019} and environments \cite{Aguilera2014,Becker2018}, and could provide a complementary alternative for all optical degenerate gases production \cite{Stellmer2013,Urvoy2019}.\medskip

We thank Leonardo Ricci for kindly providing a Rb reservoir in a critical phase of the experiment. This work was partly supported by the ``Agence Nationale pour la Recherche" (grant EOSBECMR \# ANR-18-CE91-0003-01), Laser and Photonics in Aquitaine (grant OE-TWC), Horizon 2020 QuantERA ERA-NET (grant TAIOL \# ANR-18-QUAN-00L5-02), and the Aquitaine Region (grants IASIG-3D and USOFF). Please acknowledge support for M.C. by Dstl under Contract No. DSTLX-1000097855 and T.F. by the EPSRC through the Quantum Technology Hub for Sensors \& Metrology under Grant No. EP/M013294/1.

%\bibliography{main} 


\newpage

\onecolumngrid

\begin{center}
\large{\textbf{ Supplemental material: \\ Loading and Cooling in an Optical Trap via Hyperfine Dark States}}
\end{center}

\section*{DETERMINATION OF HYPERFINE DARK STATES}

To investigate the formation of dark states, we use a 1-dimensional model of the  gray molasses \cite{Weidemller1994,Grier2013} that includes the presence of two counterpropagating beams with opposite circular polarizations ($\sigma_+/\sigma_-$ configuration), each carrying both cooler and repumper frequency, and the light shifts due to the trapping potential.

We consider the Hamiltonian as a sum of three terms: {\it (i)} the atomic Hamiltonian, taking into account only the $5 ^{2}S_{1/2}$ and $5 ^{2}P_{3/2}$ states connected by the D2 transition; {\it (ii)} the interaction of the atom with the cooler and repumper light; {\it (iii)} the interaction of the atom with the dipole trap light at 1560 nm, causing scalar and vector light shifts, especially large for to the states of the $5 ^{2}P_{3/2}$ level:

\begin{align}
H &= H_{\rm{at}} + H_{\rm{mol}} + H_{\rm{dip}} \\
H_{\rm{at}} &=\sum_{n=5S, 5P_{3/2}}\sum_{n,F,m} E \left (n,F,m \right ) |n, F, m\rangle \langle j, F, m| \\
H_{\rm{mol}} & = - \frac{\hbar}{2} \left [ \left (\Omega_Re^{-i\omega_R t}+\Omega_C e^{-i\omega_C t} \right ) \right. \nonumber\\
& \left. \times \left ( \hat{e}_+e^{ikz}+\hat{e}_-e^{-ikz} \right )\vec{a} + h.c. \right ]   \label{eq:hmol} \\
H_{\rm{dip}} &=\frac{|{\cal E}_0|^2}{2} \left[\epsilon_\mu^* \epsilon_\nu^* d_\mu\sum_k \frac{P_k (E_v - E_k)}{(E_v - E_k)^2 - (\hbar \omega_d)^2} d_\nu \right] .
\end{align}

In the atomic Hamiltonian $H_{\rm{at}}$, $E(n,F,m)$ denotes the energies of the ground and excited hyperfine manifolds, hereafter indicated with $5S$ and $5P$,  with the definition $E(5S, 1, m)=0$.

In the molasses Hamiltonian $H_{\rm{mol}}$, $\Omega_{R(C)} \equiv \Gamma \sqrt{I_{R(C)}/ \left(2 I_S \right)}$ is the repumper (cooler) Rabi frequency in terms of the saturation intensity $I_S=1.67$ mW/cm$^2$ and the excited state linewidth $\Gamma/(2\pi)=$ 6.065 MHz, $\vec{a}$ are the raising operators of atomic levels whose matrix elements are the 6-j Wigner coefficient and, finally, $\omega_{R(C)}$ the repumper (cooler) angular frequency. Each molasses beam carries the repumper and cooler frequency $\omega_R , \omega_C$, with Rabi frequencies $\Omega_C=4.2\,\Gamma, \Omega_R=1.2\,\Gamma$, corresponding to the total intensities, i.e. summed on the six beams in our experiment. Eq.~(\ref{eq:hmol}) is further simplified by neglecting the coupling of the cooler with the $F=1\rightarrow F'$ transitions, due to very large detuning ($\sim 10^3 \, \Gamma$), and likewise the coupling of the repumper with $F=2\to F'$ transitions:

\begin{align}
H_{\rm{mol}} & \simeq - \frac{\hbar}{2} \left[ \left( \Omega_Re^{-i\omega_R t}P_e \vec{a} P_1+\Omega_C e^{-i\omega_C t} P_e \vec{a} P_2 \right) \right. \nonumber\\
& \left. \times \left( \hat{e}_+e^{ikz}+\hat{e}_-e^{-ikz} \right) + h.c. \right] .
\end{align}

Finally, $H_{\rm{dip}}$ is the second-order perturbation Stark Hamiltonian for the 1560 nm light $\vec{\cal E}(t)=\frac12 {\cal E}_0 \hat\epsilon e^{-i \omega_d t} +c.c.$, with $d_\mu, \epsilon_\mu (\mu=0,\pm1)$ being the spherical components of the electric dipole operator and of the polarization vector; $E_v$ is the energy of the right ket and $k$ labels the intermediate state. 

One can apply a unitary transformation
$$U = P_1 + \exp \left[ i \left( \omega_R-\omega_C \right) t \right] P_2 +\exp \left[ i \omega_R t \right] P_e$$
where $P_1, P_2, P_e$ are the projectors on the ground lower $\{|5S,F=1,m\rangle\}$, ground upper $\{|5S,F=2,m\rangle\}$, and electronic excited $\{|5P,F',m'\rangle\}$ hyperfine levels, respectively. Under the unitary transformation $U$, the Hamiltonian is modified: (i) the time-dependence of the molasses Hamiltonian $H_{\rm{mol}}$ drops, (ii) the atomic energy levels are shifted $E'(5S,1,m)=0,\, E'(5S,2,m) = E(5S,2,m)-\hbar(\omega_R - \omega_C),\, E'(5P,F,m) = E(5P,F,m)-\hbar \omega_R$.

We set the frequency difference $\omega_R-\omega_C$ to match the hyperfine separation of a free atom and then diagonalize the Hamiltonian for different values of the position $z$.  At each position, we identify the dark states as those whose projection in the excited level $5 P$ is below an arbitrary threshold value, chosen equal to 0.1; in practice a given eigenstate $\ket\psi$ of the total Hamiltonian is considered dark if $\bra\psi P_e \ket\psi <0.1$, which is equivalent to require a scattering rate below $0.1 \, \Gamma$. We %point out 
find that, in the $\sigma_+/\sigma_-$ configuration of our 1D model, the light shifts are uniform in space, i.e. independent of $z$. Indeed at each point the polarization of the molasses lasers field is linear, with direction varying periodically in space: the polarization vector winds like a helix. As a consequence, it is expected that the number of dark states does not depend on the position $z$.

In the set of dark states, we operate a further selection by keeping only those composed of hyperfine states $\ket{F,m=\pm 1}$ and discarding those formed by $\ket{F,2}$ and $\ket{F,0}$. The former are linear superpositions of states with momenta $p\pm \hbar k_0$, that can be stable under kinetic energy evolution for sufficiently small $p$. On the other hand, the latter are  superpositions of states with momenta $p, p\pm 2\hbar k_0$, that cannot be eigenstates of the kinetic energy. For molasses times longer than the inverse kinetic energy associated to the two-photon transitions, $t_{mol}>(2 \hbar k_0^2/m)^{-1}\simeq 10\,\mu$s, the latter states become bright. 

In order to calculate the expected relative populations of the $F=1$ and $F=2$ hyperfine levels at the end of the molasses, we take into account only the dark states selected as described. On each of these states $\ket{{\rm DS}_j}$, we evaluate $p_{1,j}=\bra{{\rm DS}_j}P_1\ket{{\rm DS}_j}$ and $p_{2,j}=\bra{{\rm DS}_j}P_2\ket{{\rm DS}_j}$ and define the relative population $N_2/(N_1+N_2) = \sum_j p_{2,j}/(p_{1,j}+p_{2,j})$, implicitly assuming that all dark states are equally populated.
The purple points in Fig. 2 of the main text report this quantity in the specific case for zero power of the FORT.

\end{document}